\documentclass[preprint, 12pt]{elsarticle}

\usepackage{graphics}
\usepackage{graphicx}
\usepackage{epsfig}
 
\usepackage{amssymb}
\usepackage{amsthm}

\usepackage{lineno}

\usepackage{color}

\begin{document}

\begin{frontmatter}




\title{Statistics of finite scale local Lyapunov exponents in fully developed homogeneous isotropic turbulence}


\author{Nicola de Divitiis}

\address{"La Sapienza" University, Dipartimento di Ingegneria Meccanica e 
Aerospaziale, Via Eudossiana, 18, 00184 Rome, Italy, \\
phone: +39--0644585268, \ \ fax: +39--0644585750, \\
e-mail: n.dedivitiis@gmail.com, \ \  nicola.dedivitiis@uniroma1.it}


\begin{abstract}
The present work analyzes the statistics of finite scale local Lyapunov exponents of pairs of fluid particles trajectories in fully developed incompressible homogeneous isotropic turbulence.
According to the hypothesis of fully developed chaos, this statistics is here analyzed assuming that 
the entropy associated to the fluid kinematic state is maximum.
The distribution of the local Lyapunov exponents results to be an unsymmetrical uniform function in a proper interval of variation.
From this PDF, we determine the relationship between average and maximum Lyapunov exponents, 
and the longitudinal velocity correlation function. This link, which in turn leads to the closure of von K\'arm\'an-–Howarth and Corrsin equations, agrees with results of previous works, supporting the proposed PDF calculation, at least for the purposes of the energy cascade main effect estimation.
Furthermore, through the property that the Lyapunov vectors tend to align the direction of the maximum growth rate of trajectories distance, we obtain the link between maximum and average Lyapunov exponents in line with the previous results.
To validate the proposed theoretical results, we present different numerical simulations whose results justify the hypotheses of the present analysis. 
\end{abstract}


\end{frontmatter}
 
\newcommand{\no}{\noindent}
\newcommand{\be}{\begin{equation}}
\newcommand{\ee}{\end{equation}}
\newcommand{\bea}{\begin{eqnarray}}
\newcommand{\eea}{\end{eqnarray}}
\newcommand{\bc}{\begin{center}}
\newcommand{\ec}{\end{center}}

\newcommand{\calr}{{\cal R}}
\newcommand{\calv}{{\cal V}}

\newcommand{\bff}{\mbox{\boldmath $f$}}
\newcommand{\bfg}{\mbox{\boldmath $g$}}
\newcommand{\bfh}{\mbox{\boldmath $h$}}
\newcommand{\bfi}{\mbox{\boldmath $i$}}
\newcommand{\bfm}{\mbox{\boldmath $m$}}
\newcommand{\bfp}{\mbox{\boldmath $p$}}
\newcommand{\bfr}{\mbox{\boldmath $r$}}
\newcommand{\bfu}{\mbox{\boldmath $u$}}
\newcommand{\bfv}{\mbox{\boldmath $v$}}
\newcommand{\bfx}{\mbox{\boldmath $x$}}
\newcommand{\bfy}{\mbox{\boldmath $y$}}
\newcommand{\bfw}{\mbox{\boldmath $w$}}
\newcommand{\bfk}{\mbox{\boldmath $\kappa$}}

\newcommand{\bfA}{\mbox{\boldmath $A$}}
\newcommand{\bfD}{\mbox{\boldmath $D$}}
\newcommand{\bfI}{\mbox{\boldmath $I$}}
\newcommand{\bfL}{\mbox{\boldmath $L$}}
\newcommand{\bfM}{\mbox{\boldmath $M$}}
\newcommand{\bfS}{\mbox{\boldmath $S$}}
\newcommand{\bfT}{\mbox{\boldmath $T$}}
\newcommand{\bfU}{\mbox{\boldmath $U$}}
\newcommand{\bfX}{\mbox{\boldmath $X$}}
\newcommand{\bfY}{\mbox{\boldmath $Y$}}
\newcommand{\bfK}{\mbox{\boldmath $K$}}
\newcommand{\bfR}{\mbox{\boldmath $R$}}

\newcommand{\bfrho}{\mbox{\boldmath $\varrho$}}
\newcommand{\bfchi}{\mbox{\boldmath $\chi$}}
\newcommand{\bfphi}{\mbox{\boldmath $\phi$}}
\newcommand{\bfPhi}{\mbox{\boldmath $\Phi$}}
\newcommand{\bflambda}{\mbox{\boldmath $\lambda$}}
\newcommand{\bfell}{\mbox{\boldmath $\ell$}}
\newcommand{\bfxi}{\mbox{\boldmath $\xi$}}
\newcommand{\bfeta}{\mbox{\boldmath $\eta$}}
\newcommand{\bfLambda}{\mbox{\boldmath $\Lambda$}}
\newcommand{\bfPsi}{\mbox{\boldmath $\Psi$}}
\newcommand{\bfXi}{\mbox{\boldmath $\Xi$}}
\newcommand{\bfomega}{\mbox{\boldmath $\omega$}}
\newcommand{\bfOmega}{\mbox{\boldmath $\Omega$}}
\newcommand{\bfeps}{\mbox{\boldmath $\varepsilon$}}
\newcommand{\bfepsn}{\mbox{\boldmath $\epsilon$}}
\newcommand{\bftau}{\mbox{\boldmath $\tau$}}
\newcommand{\bfzeta}{\mbox{\boldmath $\zeta$}}
\newcommand{\bfkappa}{\mbox{\boldmath $\kappa$}}
\newcommand{\bfsigma}{\mbox{\boldmath $\sigma$}}
\newcommand{\bftheta}{\mbox{\boldmath  $\vartheta$}}
\newcommand{\bfmu}{\mbox{\boldmath  $\mu$}}
\newcommand{\itPsi}{\mbox{\it $\Psi$}}
\newcommand{\itPhi}{\mbox{\it $\Phi$}}

\newcommand{\bint}{\mbox{ \int{a}{b}} }
\newcommand{\ds}{\displaystyle}
\newcommand{\Sum}{\Large \sum}



\bigskip

\section{Introduction \label{sect1}}

In fully developed turbulence, the finite scale Lyapunov exponents of the fluid kinematic field are of paramount importance because i) describe the turbulent energy cascade phenomenon, and ii) give the fluid viscous dissipation when the length scale goes to zero. One of the characteristics of these exponents in turbulence is that their statistics is related to the instantaneous velocity field, whereas does not depend directly on the time variations of this latter. Such characteristic, which represents the crucial point of this work, is consequence of the fact that the times of variations of the velocity field, which changes according to the Navier--Stokes equations, are much greater than those of the fluid displacements which follow the fluid kinematics \cite{deDivitiis_1, deDivitiis_2}.
Specifically, whereas the velocity field is a function of slow growth of the time, the distance between particles and the local fluid deformation are unbounded quantities
which rise exponentially with the time \cite{deDivitiis_1}.
Accordingly, the fluid particles displacements statistics, being not directly linked to the time variations of the velocity field, can be considered to be the result of the current fluid act of motion.

{\color{black}
It is worth to remark that, the adoption of the finite scale Lyapunov exponents in place of the classical ones is justified by the fact that the turbulence is a complex phenomenon involving numerous length scales, each of them is characterized by different properties. The several perturbations of finite size will vary following nonlinear differential equations out of tangent space, thus the sole use of classical Lyapunov exponents is not adequate to describe the perturbations behavior associate to the different length scales and therefore the energy cascade.
}

Next, as a result of item i), the knowledge of the Lyapunov exponents statistics would lead to the closure of the equations of velocity and temperature correlations and therefore to the determination 
of kinetic energy and temperature spectra. Therefore, the distribution of such exponents is very  important for quantifying the effects of the turbulent energy cascade.

Several works dealing with the closure of the correlations equations are present in the literature, as for instance \cite{Baev,  Grebenev05, Hasselmann58, Mellor84, Millionshtchikov69, Oberlack93, Antonia2013}. Nevertheless, to the author knowledge, the influence of the finite scale Lyapunov exponents statistics on turbulent energy cascade and on the closure of the correlations equations has not received due attention.
Hence, the objective of the present work is to develop a theoretical analysis based on the aforementioned properties which leads to determine the statistics of the local finite scale Lyapunov exponent in fully developed homogeneous isotropic turbulence for incompressible fluids.
This local exponent, defined as
\bea
\ds \tilde{\lambda} \equiv \frac{d \ln \rho}{dt} = \frac{{\bf \dot{{\bfxi}}} \cdot {\bfxi}}{{\bfxi} \cdot {\bfxi}} 
\label{2}
\eea
provides the instantaneous growth rate of the distance $\rho = \sqrt{{\bfxi}\cdot {\bfxi}}$ between  two fluid particles trajectories ${\bf x}(t)$ and ${\bf y}(t)={\bf x}(t)+{\bfxi}(t)$, where $\bfxi$ is the separation vector (finite scale Lyapunov vector).
{\color{black}
This exponent is linked to the so called finite size Lyapunov exponent, {FSLE},  defined
in Ref. \cite{Vulpiani_13} in a more general framework of dynamic systems. Specifically, FSLE can be obtained in terms of $\tilde{\lambda}$  as a proper average of $\tilde{\lambda}$ over a given interval, say $T=\Sum_k (t_k-t_{k-1})$,
where $(t_k, t_{k-1})$ is the subset in which $\rho$ rises from $\rho(t_{k-1})$ to $\varrho_R \rho(t_{k-1})$,  $\varrho_R>$1 is an assigned threshold slightly greater than the unity to avoid interferences between the scales, whereas ${\bfxi}(t_k)$, k=1, 2,... are rescaled along the direction ${\bfxi}(t_k)/\vert {\bfxi}(t_k)\vert$. 
}

Because of non--smooth spatial variations of the velocity field, $\tilde{\lambda}$ 
can exhibit fluctuations of sizable amplitude with respect to its average value, 
thus $\tilde{\lambda}$ plays the role of a stochastic variable and will be 
distributed according to a certain PDF. To define the magnitude of these oscillations, average and maximum finite scale Lyapunov exponents, $\bar{\lambda}$ and $\lambda_+$ respectively, are so defined 
\bea
\begin{array}{l@{\hspace{-0.cm}}l}
\ds \bar{\lambda} \equiv \left\langle \tilde{\lambda} \right\rangle \\\\
\label{l ave}
\ds {\lambda_+} \equiv \left\langle \tilde{\lambda} \right\rangle_{\dot{\xi} \cdot \xi \ge 0} 
\label{L0}
\end{array}
\eea
where  $\ds \left\langle \circ \right\rangle$ and
$\ds \left\langle \circ \right\rangle_{\dot{\xi} \cdot \xi \ge 0}$
denote the average over the entire ensemble of $\bfxi$, and 
the average calculated on the ensemble where $\ds {\bf \dot{\bfxi}} \cdot \bfxi \ge 0$.

As the consequence of the aforementioned properties, the present analysis assumes that the kinematics of a pair of fluid particles, characterized by $\bfxi$, is much faster and statistically independent with respect to the time variations of velocity field. This property, just discussed in \cite{deDivitiis_1, deDivitiis_2} for what concerns the closure of von K\'arm\'an-–Howarth and Corrsin equations \cite{Karman38} \cite{Corrsin_1, Corrsin_2}, was previously supported by the arguments presented in Ref. \cite{Ottino90} (and references therein), where the author observes that:
i) the velocity fields ${\bf u} (t, {\bf x})$ produce chaotic trajectories also for relatively simple mathematical structure of ${\bf u} (t, {\bf x})$. ii) the flows given by ${\bf u} (t, {\bf x})$ stretch and fold continuously and rapidly causing an effective mixing of the particles trajectories.

Through the hypotheses of fully developed chaos and fluid incompressibility, we first estimate the interval of variation of $\tilde{\lambda}$, and thereafter determine the distribution function of $\tilde{\lambda}$ by maximizing the entropy associated to the fluid kinematic state. 
The maximization of such entropy is justified by the fact that the regime of fully developed turbulence corresponds to a situation of maximum chaos where the bifurcations cause a total loss of the initial condition data of $\bf x$ and $\bfxi$.
As the consequence, we show that $\tilde{\lambda}$ results to be uniformely distributed in such interval, in particular, we determine the relationship between $\bar{\lambda}$ and $\lambda_+$, resulting $\lambda_+ = 2 \bar{\lambda}$. 

A further confirmation of such link is obtained by exploiting the alignment property of $\bfxi$, following which the Lyapunov vectors tend to align  the direction of the maximum growth rate of $\rho$ \cite{Ott2002}.

{\color{black}In order to compare the proposed statistics with the average energy cascade effects,}
$\bar{\lambda}$ and $\lambda_+$ are expressed in terms of the longitudinal velocity correlation function $f = {\left\langle u_r u'_r \right\rangle }/{u^2}$
where
$u^2 = {1}/{3} \left\langle {\bf u} \cdot {\bf u}   \right\rangle$,
$u_r = {\bf u}(t, {\bf x}) \cdot {{\bf r}}/{r}$ and 
$u_r' = {\bf u}(t, {\bf x}+{\bf r}) \cdot {{\bf r}}/{r}$.
This relationship leads to the closure formulas of von K\'arm\'an--Howarth and Corrsin equations, and coincides with that just presented in Ref. \cite{deDivitiis_1} where the author 
adopts only $\lambda_+$ and $\bar{\lambda}$ without considering the distribution of the local exponents. There, the proposed closure formula leads to a value of the skewness of $\partial u_r/\partial r$ equal to -3/7, in agreement with those obtained by the several authors with direct numerical simulation of the Navier--Stokes equations (DNS) \cite{Chen92, Orszag72, Panda89}, and Large--eddy simulations (LES) \cite{Anderson99, Carati95, Kang2003}. Therefore, the here proposed statistics should be adequate for describing the distribution of $\tilde{\lambda}$, at least for what concerns the main properties of the energy cascade phenomenon.
{\color{black}
The novelty of this work with respect to the literature, and in particular with respect to  Ref. \cite{deDivitiis_1}, is represented by the statistics of $\tilde{\lambda}$ and its distribution function. This latter provides much more detailed statistical informations about dissipation and energy cascade than the analysis of Ref. \cite{deDivitiis_1} which, adopting average exponents, describes only the mean effects of energy cascade and dissipation. The present analysis provides, by means of such PDF, the estimation of all the Lyapunov exponent statistical moments. Accordingly, this study recovers, among the other things, the results of Ref. \cite{deDivitiis_1} and, in addition, gives all the statistical properties arising from this specific PDF, in particular, the deviations of such effects of dissipation and energy cascade with respect to their average values.
}

Furthermore, to justify the hypotheses of the proposed statistics, the theoretical results of this latter are compared with numerical simulations of a proper differential system representing the incompressible fluid kinematics.

\bigskip

\section{Interval of variation of $\tilde{\lambda}$ in incompressible turbulence \label{sect2}}

This section proposes an analysis for estimating the range of variations of 
$\tilde{\lambda}$ based on the hypotheses of fluid incompressibility and fully developed chaos.

The present analysis starts from the consideration that the turbulent energy cascade is related to the 
fluid particles trajectories divergence, therefore the relative fluid kinematics plays an important role in the estimation of the properties of such energy cascade \cite{deDivitiis_1, deDivitiis_2}.
The relative kinematics is expressed by finite scale Lyapunov vector $\bfxi$ which satisfies 
\bea
\begin{array}{l@{\hspace{-0.cm}}l}
\ds \dot{{\bf x} } = {\bf u} (t, {\bf x}),   \\\\
\ds \dot{{\bfxi} } = {\bf u} (t, {\bf x}+{\bfxi}) - {\bf u} (t, {\bf x})
\end{array}
\label{1}
\eea
where ${\bf u}={\bf u} (t, {\bf x})$ varies according to the Navier--Stokes equations, and ${\bf x}(t)$ and ${\bf y}(t)={\bf x}(t)+{\bfxi}(t)$ are two fluid particles trajectories. The local divergence between ${\bf x}(t)$ and ${\bf y}(t)$ is quantified by $\tilde{\lambda}$ which, due to the bifurcations of the kinematic field (kinematic bifurcations), exhibits oscillations 
of sizable amplitude with respect to its average value. These bifurcations continuously happen in those points of the space where
\bea
\ds \det \left( \nabla  {\bf u} (t, {\bf x}) \right) = 0
\label{bifurcations}
\eea 
Observe that the kinematic bifurcations defined by Eq. (\ref{bifurcations}) are not the Navier--Stokes equations bifurcations (dynamic bifurcation), but arise from these latter \cite{deDivitiis_2}. In fact, the Navier--Stokes bifurcations frequentely occur determining continuously non--smooth spatial variations of ${\bf u} (t, {\bf x})$ which in turn lead to the condition (\ref{bifurcations}) in the several points of the space.

The definition of $\tilde{\lambda}$ given by Eq. (\ref{2}) implies that ${\bfxi}$ can be locally expressed as 
\bea
\begin{array}{l@{\hspace{-0.cm}}l}
\ds {\bfxi} = {\bf Q} (t) {\bfxi}(0) \exp(\tilde{\lambda} t), 
\end{array}
\label{3}
\eea
where $\tilde{\lambda}$ plays the role of the stochastic variable and
${\bf Q} (t)$ is a proper orthogonal matrix providing the orientation of ${\bfxi}$ with respect to the
inertial frame $\mathcal R$. Accordingly, ${\bf \dot{\bfxi}}$ is
\bea
\begin{array}{l@{\hspace{-0.cm}}l}
\ds {\bf \dot{\bfxi}} = \tilde{\lambda} \ {\bfxi} + \bfomega \times {\bfxi}
\end{array}
\label{3'}
\eea
in which $\bfomega$ defines the angular velocity of ${\bfxi}$ with respect to $\cal R$.

For an assigned velocity field, finite scale Lyapunov exponents and vectors are formally calculated 
with  Eq. (\ref{2}) through the following orthogonalization  
 procedure:
1) the maximal local Lyapunov exponent, say  $\tilde{\lambda}_1$, is first obtained by choosing the direction 
${\bf e}_1(t) \equiv \bfxi_1/\vert \bfxi_1 \vert$ which maximizes $\tilde{\lambda}$ in Eq. (\ref{2}), b) the second exponent $\tilde{\lambda}_2 \leq \tilde{\lambda}_1$ is calculated by selecting the direction ${\bf e}_2(t) \equiv \bfxi_2/\vert \bfxi_2 \vert$ in the subspace (plane) orthogonal to $\bfxi_1$ which maximizes $\tilde{\lambda}$, c) finally, the third one  $\tilde{\lambda}_3 \leq \tilde{\lambda}_2 \leq \tilde{\lambda}_1$ corresponds to the direction 
${\bf e}_3 \equiv \bfxi_3/\vert \bfxi_3 \vert$ normal to both ${\bf e}_1$ and ${\bf e}_2$, where  $\vert \bfxi_1 \vert$=$\vert \bfxi_2 \vert$=$\vert \bfxi_3 \vert$.
The so obtained vectors system $E \equiv ({\bf e}_1, {\bf e}_2, {\bf e}_3)$ defines a rigid space which moves with respect to $\cal R$ with a given angular velocity $\bfomega_E$ depending on the local fluid motion.
Therefore, ${\bfxi}_k$ and $\tilde{\lambda}_k$, $k=1, 2, 3$, are locally expressed by
\bea
\begin{array}{l@{\hspace{-0.cm}}l}
\ds {\bfxi}_k = {\bf Q} (t) {\bfxi}_k(0) \exp(\tilde{\lambda}_k t), \\\\ 
\ds {\bf \dot{\bfxi}}_k = \tilde{\lambda_k}{\bfxi}_k + \bfomega_E \times {\bfxi}_k, 
\end{array}
 \ \ \ \ \ \ \ \ \ k = 1, 2, 3.
\label{4}
\eea
The classical local Lyapunov exponents $\tilde{\Lambda}_k$ are defined for ${\bfxi}_k \rightarrow 0$, $k=1, 2, 3$. 

Now, in order to estimate the set of variations of $\tilde{\lambda}$, observe that,
due to fluid incompressibility, $\nabla \cdot {\bf u} \equiv 0$, and the classical local exponents obey to the following condition
\bea
\begin{array}{l@{\hspace{-0.cm}}l}
\ds \tilde{\Lambda}_1+ \tilde{\Lambda}_2 + \tilde{\Lambda}_3=0, 
\end{array}
\label{5}
\eea
In general, Eq.(\ref{5}) does not hold for finite scale Lyapunov vectors.
Nevertheless, Eq. (\ref{5}) is valid for those finite scale vectors 
for which the volume ${\bfxi}_1 \times {\bfxi}_2 \cdot {\bfxi}_3$ 
is locally preserved, therefore, without lack of generality, such these exponents can be written in the form
\bea
\begin{array}{l@{\hspace{-0.cm}}l}
\ds \tilde{\lambda}_1 = \lambda_m \cos\left( \varepsilon \right), \\\\
\ds \tilde{\lambda}_2 = \lambda_m \cos\left( \varepsilon + \frac{2}{3} \pi \right), \\\\
\ds \tilde{\lambda}_3 = \lambda_m \cos\left( \varepsilon + \frac{4}{3} \pi\right)
\end{array}
\label{iso lambda}
\eea
where $\varepsilon$ and $\lambda_m$ are variables depending on
the current act of motion, and
\bea
\begin{array}{l@{\hspace{-0.cm}}l}
\ds \tilde{\lambda}_1+ \tilde{\lambda}_2 +\tilde{\lambda}_3=0, \ \ \tilde{\lambda}_1 \ge \tilde{\lambda}_2 \ge \tilde{\lambda}_3.
\end{array}
\label{6}
\eea
Now, following the hypothesis of fully developed chaos, it is reasonable that $\tilde{\lambda}$ ranges in the set $(\lambda_0, \lambda_S)$, where $\lambda_S>0$ assumes its maximum value
compatible with Eq. (\ref{iso lambda}), and $\lambda_0$ is consequentely calculated.
This implies that 
$\ds \lambda_0 = - \lambda_S/2$, that is
\bea
\begin{array}{l@{\hspace{-0.cm}}l}
\ds \tilde{\lambda} \in \left( -\frac{{\lambda_S}}{2}, {\lambda_S}\right), \\\\
\ds \lambda_S = \sup\left\lbrace \lambda_m \right\rbrace  
\end{array}
\eea
It is worth to remark that $\lambda_S$ depends on the instantaneous velocity field which in turn is the result of time evolution of the fluid motion starting from the initial condition following the Navier--Stokes equations. Hence, at the current time, $\lambda_S$ assumes values related to viscous dissipation and kinetic energy both consequence of the fluid motion.

\bigskip

\section{Incompressible fully developed turbulence \label{sect3}}

This section studies the distribution functions $P= P(t, {\bf x}, {\bfxi})$ and $P_\lambda= P_\lambda(t, \tilde{\lambda})$ in incompressible fully developed turbulence, where $P(t, {\bf x}, {\bfxi})$ and $ P_\lambda(t, \tilde{\lambda})$, are, respectively, the distribution functions of fluid particles position and Lyapunov vector, and of local Lyapunov exponent.
In particular, we will show that the proposed statistics leads to the following relations
\bea
\begin{array}{l@{\hspace{-0.cm}}l}
\ds \left\langle  \tilde{\lambda} \right\rangle  =\frac{\lambda_+}{2},
\end{array}
\label{eq0}
\eea
\bea
\begin{array}{l@{\hspace{-0.cm}}l}
\ds \left\langle  \tilde{\lambda}^2 \right\rangle =  \lambda_+^2 
\end{array}
\eea
and to the link between $\lambda_+$ and the longitudinal velocity correlation function.

According to Eq. (\ref{2}), $P_\lambda$ depends on $P$, and is expressed in terms of this latter through the Frobenius Perron equation
\bea
\begin{array}{l@{\hspace{-0.cm}}l}
\ds P_\lambda(t, \tilde{\lambda}) = \int_{\bf x} \int_{\bf \xi} P(t, {\bf x}, {\bfxi}) \ 
\delta\left( \tilde{\lambda} - \frac{{\bf \dot{{\bfxi}}} \cdot {\bfxi}}{{\bfxi} \cdot {\bfxi}} \right) \ d {\bf x} \ d {\bfxi},
\end{array}
\label{fb}
\eea
where $\delta$ is the Dirac's delta.
Hence, the distribution function $P$ is first studied. 
This PDF changes with the time according to the Liouville theorem \cite{Nicolis95} associated to Eqs. (\ref{2}). This theorem, arising from the following relation 
\bea
\begin{array}{l@{\hspace{-0.cm}}l}
\ds \int_{\bf x} \int_{\bf \xi}   P \ d {\bf x} \ d {\bfxi}  = 1, \ \ \ \forall t>0,
\end{array}
\label{3 L}
\eea
and from Eqs. (\ref{2}), provides the evolution equation of $P$ \cite{Nicolis95}
\bea
\begin{array}{l@{\hspace{-0.cm}}l}
\ds \frac{\partial P}{ \partial t} +
 \nabla_{\bf x} \cdot \left( P {\bf \dot{{\bf x}}}\right)+
 \nabla_{\bf \xi} \cdot \left( P {\bf \dot{{\bfxi}}}\right) = 0
\end{array}
\label{liouville L}
\eea
where $\nabla_{\bf x} \cdot \left( \circ \right)$ and $\nabla_{\bf \xi} \cdot \left( \circ \right)$
denote the divergence of $(\circ)$ defined in the spaces $\left\lbrace \bf x\right\rbrace$ and 
$\left\lbrace \bfxi \right\rbrace$ respectively, and $d {\bf x}$ and $d {\bfxi}$ are the elemental volumes in the corresponding spaces.

Taking into account Eq. (\ref{3 L}), and that the homogeneous isotropic turbulence is defined for unbounded fluid domains, $P$ will satisfy the following boundary condition 
\bea
\begin{array}{l@{\hspace{-0.cm}}l}
\ds P = 0, \ \forall ({\bf x}, {\bfxi}) \in \partial \left\lbrace \left\lbrace {\bf x} \right\rbrace \times \left\lbrace {\bfxi} \right\rbrace \right\rbrace \equiv \partial \left\lbrace {\bf x} \right\rbrace 
\bigcup \partial \left\lbrace {\bfxi} \right\rbrace 
\end{array}
\label{bc}
\eea
Accordingly, the statistical average of an integrable function of $\bf x$ and $\bfxi$, say $\zeta$, is
calculated in terms of $P$ 
\bea
\begin{array}{l@{\hspace{-0.cm}}l}
\ds \left\langle {\zeta} \right\rangle = 
 \int_{\bf x} \int_{\bf \xi}  P \ \zeta \ d  {\bf x} \ d {\bfxi},
\end{array}
\label{ave L}
\eea 
In particular, average and maximum finite scale Lyapunov exponents are  
\bea
\begin{array}{l@{\hspace{-0.cm}}l}
\ds \left\langle \tilde{\lambda} \right\rangle =
 \int_{\bf x} \int_{\bf \xi} P(t, {\bf x}, {\bfxi}) \
\frac{{\bf \dot{{\bfxi}}} \cdot {\bfxi}}{{\bfxi} \cdot {\bfxi}} \ d {\bf x} \ d {\bfxi} 
\equiv
\int_\lambda P_\lambda(t, \tilde{\lambda}) \tilde{\lambda} \ d \tilde{\lambda},
\end{array}
\label{l ave}
\eea
\bea
\begin{array}{l@{\hspace{-0.cm}}l}
\ds {\lambda_+} =
\frac{ \ds \int_{\bf x} \int_{\dot{\xi} \cdot \xi \ge 0} P(t, {\bf x}, {\bfxi}) \ 
\frac{{\bf \dot{{\bfxi}}} \cdot {\bfxi}}{{\bfxi} \cdot {\bfxi}} \ d {\bf x} \ d {\bfxi}} 
{ \ds \int_{\bf x} \int_{\dot{\xi} \cdot \xi \ge 0} P(t, {\bf x}, {\bfxi}) 
\  d {\bf x} d {\bfxi} } 
=
\frac{ \ds \int_{\tilde{\lambda} \ge 0}  P_\lambda(t, \tilde{\lambda}) \ 
\tilde{\lambda} \ d \tilde{\lambda} } 
{ \ds \int_{\tilde{\lambda} \ge 0} P_\lambda(t, \tilde{\lambda}) \ d  \tilde{\lambda} } 
\end{array}
\label{l +}
\eea

{\bf Remark}: 
It is very important to observe that the Liouville theorem in the form of Eq. (\ref{liouville L}) holds also when the variable velocity field is replaced with the same field frozen at the current time. This is due to the hypothesis that the times of variations of the velocity field, which changes according to the Navier–Stokes equations, are much larger than those associated with  $\bf x$ and $\bfxi$  whose times of variations are of the order of $1/\tilde{\lambda}$. Following such hypothesis, the statistics of  $\bf x$ and $\bfxi$  does not depend on the time variations of $\bf u$ and can be considered to be the result of the instantaneous fluid act of motion.

\bigskip

\subsection{\bf Distribution function of ${\bf x}$ and $\bfxi$}

For sake of our convenience, we introduce the quantity 
${\bf y} \equiv ({\bf x}, {\bfxi}) \in {\cal A} \subset \left\lbrace{\bf y} \right\rbrace$ which defines the relative kinematics, being $\left\lbrace{\bf y} \right\rbrace  \equiv \left\lbrace{\bf x} \right\rbrace \times \left\lbrace{\bfxi} \right\rbrace$, thus the Liouville theorem reads as
\bea
\begin{array}{l@{\hspace{-0.cm}}l}
\ds \frac{\partial P}{ \partial t} +
 \nabla_{\bf y} \cdot \left( P {\bf f} \right) = 0
\end{array}
\label{liouville II}
\eea
where 
\bea
\begin{array}{l@{\hspace{-0.cm}}l}
\ds \dot{\bf y} = {\bf f}({\bf y})  \equiv \left(  {\bf u}(t,  {\bf x}), \  {\bf u}(t, {\bf x} +{\bfxi} )-{\bf u}(t, {\bf x})\right) 
\end{array}
\eea
Now, the entropy $\cal H$ associated to ${\bf y}$ is defined as
\bea
{\cal H}(P) = - \int_{\bf x} \int_{\xi} P \ln P \ d {\bf x} \ d {\bfxi} \equiv
- \int_{\bf y}  P \ln P \ d {\bf y} 
\eea
Due to fluid incompressibility, $\ds \nabla_{\bf y} \cdot {\bf f} \equiv 0$, therefore
from the Liouville theorem the entropy rate identically vanishes  
\bea
\ds \frac{d {\cal H}}{dt} =  \int_{\bf y} P \ \nabla_{\bf y} \cdot {\bf f} \ d {\bf y} \equiv 0  
\eea
In order to estimate the steady distribution of $\bf y$, observe that the fully developed turbulence corresponds to a situation of maximum chaos where the kinematic bifurcations cause a total loss of the initial condition data of $\bf y$.
Accordingly, it is reasonable that ${\cal H}$ assumes its maximum value compatible with 
Eqs. (\ref{3 L}) and (\ref{liouville II}) with $\partial P/\partial t=0$. This corresponds to the following variational problem
\bea
\begin{array}{l@{\hspace{-0.cm}}l}
\ds J =\int_{{\bf y}}  {\cal L} \ d {\bf y} = \max, \\\\
\end{array}
\eea
in which 
\bea
\begin{array}{l@{\hspace{-0.cm}}l}
{\cal L}
 (P, \nabla_{\bf y} P) =  - P \ln  P 
+  \eta P
+  \chi  \nabla_{\bf y} \cdot (P  \ {\bf f}) 
\end{array}
\label{lagrangian function}
\eea
is the lagrangian of the problem, and $\eta$ and $\chi=\chi({\bf y})$ are the Lagrange multipliers associated to the conditions (\ref{3 L}) and (\ref{liouville II}), respectively.
The maximum of $J$ is then obtained as steady condition for $J$ ($\delta J =0$)
through the variational calculus and this leads to the Euler--Lagrange
equation
\bea
\begin{array}{l@{\hspace{-0.cm}}l}
\ds \frac{\partial {\cal L}}{\partial P} - \nabla_{\bf y} \cdot \left( \frac{\partial {\cal L}}{\partial \nabla_{\bf y} P}\right) =0
\end{array}
\label{Euler Lagrange}
\eea 
whose solutions are searched for substituting Eq. (\ref{lagrangian function})  in Eq. (\ref{Euler Lagrange})
\bea
\begin{array}{l@{\hspace{-0.cm}}l}
\ds  P = A \exp\left( \nabla_{\bf y} \chi \cdot {\bf f} \right)
\end{array}
\label{PDF}
\eea
where $A$ is the normalization constant related to $\eta$ whose value is calculated with Eq. (\ref{3 L}), whereas $\chi(\bf y)$ is obtained through the Liouville equation. This latter gives
\bea
\begin{array}{l@{\hspace{-0.cm}}l}
\ds  \nabla_{\bf y} \cdot \left( \left( {\bf f} \otimes {\bf f}\right) \nabla_{\bf y} \chi \right)  =0
\end{array}
\label{x}
\eea 
in which $( \circ \otimes \circ)$ denotes the dyadic product between vectors. Integrating Eq. (\ref{x}), we obtain
\bea
\begin{array}{l@{\hspace{-0.cm}}l}
\ds   \left( {\bf f} \otimes {\bf f}\right) \nabla_{\bf y} \chi = 
{\bfPhi}({\bf y})
\end{array}
\label{xx}
\eea 
where ${\bfPhi}$ is an arbitrary solenoidal vector field. Now, the determinant of
$\left( {\bf f} \otimes {\bf f}\right)$ identically vanishes, thus Eq. (\ref{xx})
admits solutions only for ${\bfPhi} \equiv 0$. Moreover, as $\left( {\bf f} \otimes {\bf f}\right)$
exhibits minimum rank, Eq. (\ref{xx}) has only the trivial solution
\bea
\ds \nabla_{\bf y} \chi =0, \ \forall {\bf y} \in  {\cal A} \subset  \left\lbrace {\bf y} \right\rbrace 
\eea 
i.e. $\bf y$ is uniformely distributed on ${\cal A} \subset  \left\lbrace {\bf y} \right\rbrace$, being
\bea
P= 
\left\lbrace 
\begin{array}{l@{\hspace{-0.cm}}l}
\ds \frac{1}{m\left\lbrace {\cal A} \right\rbrace }, \ \forall {\bf y} \in  {\cal A} \subset  \left\lbrace {\bf y} \right\rbrace \\\\
\ds 0 \ \ \mbox{elsewhere} 
\end{array}\right. 
\label{P_A}
\eea
in which $m\left\lbrace \circ \right\rbrace$ indicates the measure of the set $\circ$.

\bigskip

\subsection{\bf Lyapunov exponents distribution function}

In order to estimate $P_\lambda$, the following should be noted:

If $\tilde{\lambda}$ is given, Eq. (2) corresponds to a hypersurface $\ds \Sigma_{\tilde{\lambda}}$ of the space $\left\lbrace \bf y \right\rbrace$ whose equation reads as
\bea
\begin{array}{l@{\hspace{-0.cm}}l}
\ds \Sigma_{\tilde{\lambda}}: \ds G({\bf y}; \tilde{\lambda}) \equiv  - \tilde{\lambda} + \frac{\dot{\bfxi}({\bf x}, {\bfxi})\cdot{\bfxi}}{{\bfxi}\cdot{\bfxi}} =0
\end{array}
\label{sigma}
\eea
When $\tilde{\lambda}$ varies,  $\Sigma_{\tilde{\lambda}}$ and its points ${\bf y}_* \equiv ({\bf x}_*, {\bfxi}_*)$ will change according to Eq. (\ref{sigma})
\bea
\ds \frac{d {\bf y}_*}{d \tilde{\lambda}} = \frac{\bf n}{\vert \nabla_{\bf y} G \vert} +\beta {\bftau}
\label{dec}
\eea
where ${\bf n}\equiv \nabla_{\bf y} G/\vert \nabla_{\bf y} G \vert$ and ${\bftau}$ are, respectively, local normal and tangent unit vectors to $\Sigma_{\tilde{\lambda}}$ both 
calculated in ${\bf y}_*$, being $\beta {\bftau}$ not determined. 
On the other hand, ${d {\bf y}_*}/{d \tilde{\lambda}}$ can be determined
considering that $\bf x_*$ does really not depend on $\tilde{\lambda}$,
whereas the variations of $\bfxi_*$ are related to those of $\tilde{\lambda}$ by means of Eqs. (\ref{3}) and (\ref{3'}). Thus, ${d {\bf y}_*}/{d \tilde{\lambda}}$ is locally expressed as
\bea
\ds \frac{d {\bf y}_*}{d \tilde{\lambda}} \equiv \left({\bf 0}, \frac{d {\bfxi}_*}{d \tilde{\lambda}}  \right) = \left( {\bf 0}, \frac{t}{\tilde{\lambda}} \left( \dot{\bfxi} -\omega \times {\bfxi} \right) \right)
\eea
As $\dot{\bfxi}$ is solenoidal, 
the surface integrals of $d{\bf y}_*/d \tilde{\lambda} \cdot {\bf n}$ over arbitrary hypersurfaces $\Sigma_{\tilde{\lambda}}$ ( flow of $d{\bf y}_* /d \tilde{\lambda}$ through $\Sigma_{\tilde{\lambda}}$ ) assume the same value which does not depend on $\tilde{\lambda}$, i.e.
\bea
\begin{array}{l@{\hspace{-0.cm}}l}
\ds \forall  \tilde{\lambda}_1, \tilde{\lambda}_2 \in \left(-\frac{\lambda_S}{2}, \lambda_S \right) \\\\
\ds \int_{\Sigma_{\tilde{\lambda}_1}} \left( \frac{d {\bf y}_*}{d \tilde{\lambda}} \right) \cdot {\bf n}  d \sigma =
\int_{\Sigma_{\tilde{\lambda}_2}} \left( \frac{d {\bf y}_*}{d \tilde{\lambda}} \right) \cdot {\bf n}  d \sigma,
\end{array}
\eea
therefore, from Eq. (\ref{dec}), we obtain
\bea
\begin{array}{l@{\hspace{-0.cm}}l}
\ds \forall  \tilde{\lambda}_1, \tilde{\lambda}_2 \in \left(-\frac{\lambda_S}{2}, \lambda_S \right) \\\\
\ds \int_{\Sigma_{\tilde{\lambda}_1}} \frac{d \sigma}{ \vert \nabla_{\bf y} G \vert}  =
\int_{\Sigma_{\tilde{\lambda}_2}} \frac{d \sigma}{ \vert \nabla_{\bf y} G \vert}, 
\end{array}
\label{s1=s2}
\eea

Now, the distribution function of $\tilde{\lambda}$ is calculated substituting Eq. (\ref{P_A}) in  the Frobenius--Perron equation (\ref{fb}) and considering that $\bf y$ is uniformely distributed 
on $\cal A$
\bea
\begin{array}{l@{\hspace{-0.cm}}l}
\ds P_\lambda(\tilde{\lambda}) = \int_{\bf y} P({\bf y}) 
\ \delta(G({\bf y}; \tilde{\lambda} )) d {\bf y} =
\frac{1}{m({\cal A})} \int_{\cal A}  \ \delta(G({\bf y}; \tilde{\lambda} )) d {\bf y}  \\\\
\ds = \frac{1}{m({\cal A})} \int_{\Sigma_\lambda}  \frac{d \sigma}{\vert \nabla_{\bf y} G \vert} 
\end{array}
\eea
where the surface integral over $\Sigma_\lambda \subset {\cal A}$, described by 
$G({\bf y}; \tilde{\lambda})=0$, is formally calculated according to the Minkowski measure theory \cite{Federer69}. Taking into account Eq. (\ref{s1=s2}) and that 
$\tilde{\lambda} \in \left( -{\lambda}_S/2, {\lambda}_S \right)$,  $\tilde{\lambda}$ results to be uniformely distributed in $\left( -{\lambda}_S/2, {\lambda}_S\right)$, being
\bea
\ds P_\lambda = 
\left\lbrace 
\begin{array}{l@{\hspace{-0.cm}}l}
\ds \frac{2}{3}\frac{1}{{\lambda}_S}, \ \ \mbox{if} \  \tilde{\lambda} \in \left( -\frac{{\lambda}_S}{2}, {\lambda}_S\right)  \\\\
\ds 0 \ \ \mbox{elsewhere} 
\end{array}\right. 
\label{Pl}
\eea
Hence, $\bar{\lambda}$ and $\lambda_+$ are calculated in terms of $P_\lambda$ with
Eqs. (\ref{l ave}) and (\ref{l +})  
\bea
\begin{array}{l@{\hspace{-0.cm}}l}
\ds \bar{\lambda} \equiv \left\langle \tilde{\lambda} \right\rangle   = \frac{{\lambda}_S}{4} >0, \\\\
\ds \lambda_+ \equiv \left\langle \tilde{\lambda}  \right\rangle_{\dot{\xi} \cdot \xi \ge 0} = \frac{{\lambda}_S}{2} = 2 \left\langle \tilde{\lambda}\right\rangle
\end{array}
\label{back}
\eea
Next, it is usefull to calculate the mean square $\langle \tilde{\lambda}^2 \rangle$  
\bea
\begin{array}{l@{\hspace{-0.cm}}l}
\ds \left\langle  \tilde{\lambda}^2 \right\rangle = \int_{-{\lambda}_S/2}^{{\lambda}_S}
P_\lambda \tilde{\lambda}^2 \ d \tilde{\lambda} = \lambda_+^2.
\end{array}
\label{lambda^2}
\eea
According to Eq. (\ref{lambda^2}), the mean square of $\tilde{\lambda}$ equals the square of the average of $\tilde{\lambda}$ calculated for $ {\bf \dot{\bfxi}} \cdot {\bfxi} \ge 0$, and the standard deviation is proportional to $\langle  \tilde{\lambda} \rangle$
\bea
\begin{array}{l@{\hspace{-0.cm}}l}
\ds \sigma = \sqrt{\left\langle  \tilde{\lambda}^2 \right\rangle -\left\langle  \tilde{\lambda} \right\rangle^2 } = \sqrt{3} \left\langle  \tilde{\lambda} \right\rangle = \frac{\sqrt{3}}{4} \lambda_S
\end{array}
\eea

It is worth to remark that the obtained distribution (\ref{Pl}) is the result of three effects, of which two are in competition with each other. The first one of these, due to the Lyapunov vectors tendency to align the maximum growth rate direction of $\rho$, is responsible for variation intervals in which $\tilde{\lambda}>$0, producing trajectories instability. The second one, related to the fluid incompressibility, acts in opposite sense preserving the volume, determining regions where $\tilde{\lambda}<$0.
The third element is the chaotic regime, here given by imposing $\cal H$ =max, which causes a continuous distribution of $\tilde{\lambda} \in (-\lambda_S/2, \lambda_S)$.

\bigskip

\section{Analysis through alignment property of $\bfxi$ \label{sect3 a}}

One reasonable confirmation of the previous results is here given by exploiting the alignment property of $\bfxi$, following which $\bfxi$ tends to align to the direction of the maximum growth rate of $\rho$ \cite{Ott2002}, and the fluid incompressibility. This leads to an alternative way to achieve Eq. (\ref{eq0}).
Accordingly, $\lambda_+$ is now calculated adopting a proper PDF 
$P_+= P_+(t, {\bf x}, {\bfxi})$ obtained as projection of $P(t, {\bf x}, {\bfxi})$ at the time $t+ \tau$, where ${\bf x}$ is considered to be constant and $\tau$ is the Lyapunov time defined by  
\bea
\begin{array}{l@{\hspace{-0.cm}}l}
\ds \frac{d \ln \rho}{dt} = \frac{\left\langle \ln \rho \right\rangle - \ln \rho }{\tau}
\end{array}
\label{lnro_t}
\eea 
After the time $\tau$, the alignment tendency of $\bfxi$ provides that mostly all the Lyapunov vectors calculated at ${\bf x}=$const will be such that ${\bf \dot{\bfxi}} \cdot {\bfxi} \ge 0$.
The vectors lying in subspaces of $\left\lbrace {\bfxi} \right\rbrace$ orthogonal to the maximum rising rate direction of $\rho$ which do not follow such alignment form a null measure set in $\left\lbrace {\bf y} \right\rbrace$, thus $P_+$ is a distribution function which satisfies
\bea
\begin{array}{l@{\hspace{-0.cm}}l}
\forall ({\bf x}, {\bfxi}) \in \left\lbrace {\bf x}\right\rbrace \times \left\lbrace {\bfxi} \right\rbrace \ \ \vert \ \ \ \ {\bf \dot{\bfxi}} \cdot {\bfxi} \ge 0 \ \  P_+(t, {\bf x}, {\bfxi}) > 0, \\\\
\forall ({\bf x}, {\bfxi}) \in \left\lbrace {\bf x}\right\rbrace \times \left\lbrace {\bfxi} \right\rbrace \ \ \vert \ \ \ \ {\bf \dot{\bfxi}} \cdot {\bfxi} < 0 \ \  P_+(t, {\bf x}, {\bfxi}) = 0
\end{array}
\label{prop1}
\eea
and is expressed in function of $P$ and $\tau$
\bea
\begin{array}{l@{\hspace{-0.cm}}l}
P_+ \left( t, {\bf x}, {\bfxi}\right)= P\left( t, {\bf x}, {\bfxi} + {\bf \dot{\bfxi}} \tau 
 + O(\tau^2) \right) =
P + \nabla_\xi P \cdot {\bf \dot{\bfxi}} \tau + O(\tau^2)
\end{array}
\eea
where 
$P=P\left( t, {\bf x}, {\bfxi} \right)$ and 
$\nabla_\xi P=\nabla_\xi P\left( t, {\bf x}, {\bfxi} \right)$.
Neglecting the higher order terms, $P_+$ is calculated as
\bea
\begin{array}{l@{\hspace{-0.cm}}l}
P_+ = P + \nabla_\xi P \cdot {\bf \dot{\bfxi}} \tau 
\end{array}
\label{P_+}
\eea
This PDF identically satisfies Eqs. (\ref{3 L}).
In fact, 
the integral over  
$\left\lbrace {\bf x} \right\rbrace \times \left\lbrace{\bfxi} \right\rbrace$ of the first
term at the R.H.S. of Eq. (\ref{P_+}) is equal to one, whereas the integral of the second one
can be reduced to a proper surface integral of $P$ over 
$\partial \left\lbrace {\bfxi} \right\rbrace$ where $P \equiv 0$ through Green's identity, thus this  identically vanishes.
Furthermore $P_+$ exhibits the same entropy of $P$ at least of higher order terms, which in turn
does not vary with the time due to the fluid incompressibility
\bea
\begin{array}{l@{\hspace{-0.cm}}l}
\ds {\cal H}(P_+) = - \int_{\bf x} \int_{\xi} P_+ \ln P_+ d {\bf x} \ d {\bfxi}  \\\\
\ds ={\cal H}(P) - 
\tau \int_{\bf x} \int_{\xi} \nabla_\xi P \cdot {\bf \dot{\bfxi}} \left(1 + \ln P \right) d {\bf x} \ d {\bfxi} + O(\tau^2) 
\end{array}
\label{entropy_+}
\eea
The second addend at the R.H.S. of Eq. (\ref{entropy_+}) identically vanishes as it can be reduced to be a surface integral of $P$ over $\partial \left\lbrace {\bfxi} \right\rbrace$ where $P \equiv 0$. Therefore
\bea
\begin{array}{l@{\hspace{-0.cm}}l}
\ds {\cal H}(P_+) = {\cal H}(P) + O(\tau^2) 
\end{array}
\label{entropy_++}
\eea
At least of higher order terms, $P_+$ maintains the same level of informations of $P$, thus it is adequate to estimate $\lambda_+$.
Accordingly, this latter is calculated as
\bea
\begin{array}{l@{\hspace{-0.cm}}l}
\ds {\lambda_+} =
\ds \int_{\bf x} \int_{\xi} P_+(t, {\bf x}, {\bfxi}) \
\frac{\bf {\dot{{\bfxi}}} \cdot {\bfxi}}{{\bfxi} \cdot {\bfxi}} \ d {\bf x} \ d {\bfxi}
\end{array}
\label{l max 2} 
\eea
Substituting Eq. (\ref{P_+}) and (\ref{lnro_t}) into Eq. (\ref{l max 2}), 
we have
\bea
\begin{array}{l@{\hspace{-0.cm}}l}
\ds {\lambda_+} = \left\langle \tilde{\lambda} \right\rangle -
\ds \int_{\bf x} \int_{\xi} 
\nabla_{\xi} P \cdot {\bf \dot{\bfxi}} \
\left( \ln \rho - \left\langle \ln \rho \right\rangle \right) 
\ d {\bf x} \ d {\bfxi}
\end{array}
\label{ap1}
\eea
Integrating by parts the second addend and taking into account the fluid incompressibility and the boundary conditions (\ref{bc}), we obtain
\bea
\ds \int_{\bf x} \int_{\xi} 
\nabla_{\xi} P \cdot {\bf \dot{\bfxi}}
\left( \ln \rho - \left\langle \ln \rho \right\rangle \right) 
\ d {\bf x} \ d {\bfxi} = -
\ds \int_{\bf x} \int_{\xi} P \ 
\frac{{\bf \dot{{\bfxi}}} \cdot {\bfxi}}{{\bfxi} \cdot {\bfxi}} \ d {\bf x} \ d {\bfxi} \equiv - \left\langle \tilde{\lambda} \right\rangle
\eea
Hence
\bea
\begin{array}{l@{\hspace{-0.cm}}l}
\ds \lambda_+  = 2 \left\langle \tilde{\lambda} \right\rangle
\end{array} 
\eea
in agreement with previous results.

\bigskip

\section{Lyapunov exponents in terms of longitudinal velocity correlation \label{sect4}}

The results of the previous analysis allow to achieve the link between $\lambda_+$ and the longitudinal velocity correlation function. 
In fact, the standard deviation of longitudinal velocity difference directly depends on $u^2$ and $f$ according to
\bea
\begin{array}{l@{\hspace{-0.cm}}l}
\ds  \left\langle (u_r'-u_r)^2 \right\rangle = 2 u^2 \left( 1- f(r) \right) 
\end{array} 
\eea 
On the other hand, the Lyapunov theory gives the longitudinal velocity difference in terms of $\tilde{\lambda}$
\bea
\begin{array}{l@{\hspace{-0.cm}}l}
\ds u_r'-u_r = 
\tilde{\lambda} \left( {\bfxi} \cdot \frac{\bfxi}{\vert \bfxi \vert}\right)_{\ds {\bfxi}={\bf r}}
\end{array}
\eea
Taking into account the isotropy and Eq. (\ref{lambda^2}), $\langle (u_r'-u_r)^2 \rangle$ 
reads as 
\bea
\begin{array}{l@{\hspace{-0.cm}}l}
\ds \left\langle (u_r'-u_r)^2 \right\rangle \equiv \left\langle  \dot{r}^2 \right\rangle  =
 \left\langle \tilde{\lambda}^2  \right\rangle r^2 = \lambda_+^2 r^2,
\end{array} 
\eea
thus,  $\lambda_+= \lambda_+(r)$,  $\bar{\lambda}=\bar{\lambda}(r)$ are expressed in function of the finite scale $r$ by means of the longitudinal velocity correlation $f$
\bea
\begin{array}{l@{\hspace{-0.cm}}l}
\ds \bar{\lambda} (r)=\frac{\lambda_+(r)}{2} = \frac{u}{r}\sqrt{\frac{1-f(r)}{2}} 
\end{array} 
\label{eq main'}
\eea 
Equation (\ref{eq main'}) coincides with that proposed in Ref. \cite{deDivitiis_1} which leads to the closure formulas of the von K\'arm\'an and Corrsin equations.
Unlike Ref. \cite{deDivitiis_1},  Eq. (\ref{eq main'}) is here achieved exploiting the shape of the distribution (\ref{Pl}).
\begin{table}[h]
\centering
\caption{Comparison of the results: Skewness of $\partial u_r/ \partial r$ at diverse Taylor--scale Reynolds number $R_T \equiv u \lambda_T/\nu$ following different authors.}
\vspace{2. mm}
\begin{tabular}{cccc} 
\hline
\hline
Reference   &  Simulation  &  $R_T$  & $H_3(0)$  \\
\hline 
Present analysis      &  -  &  -    & -3/7 = -0.428...  \\
\cite{Chen92}       & DNS &    202 & -0.44   \\ 
\cite{Orszag72}      & DNS &    45 & -0.47    \\
\cite{Panda89}      & DNS &    64 & -0.40    \\
\cite{Anderson99}   & LES &    $<$ 71 &  -0.40          \\ 
\cite{Carati95}     & LES &   $\infty$     &  -0.40 \\  
\cite{Kang2003}     & LES &  720        &    -0.42      \\ 
\hline
\hline
 \end{tabular}
\label{table1}
\end{table} 
Ref. \cite{deDivitiis_1} shows that Eq. (\ref{eq main'}) provides a value of the skewness of 
$\partial u_r/\partial r$ equal to $\langle (\partial u_r/\partial r)^3 \rangle/\langle (\partial u_r/\partial r)^2 \rangle^{3/2} =$ $-3/7=-0.428...$ (see the appendix), in good agreement with the results obtained by the several authors with direct numerical simulation of the Navier--Stokes equations (DNS) \cite{Chen92, Orszag72, Panda89} ($-0.47 \div -0.40$), and Large--eddy simulations (LES) \cite{Anderson99, Carati95, Kang2003} ($-0.42 \div -0.40$). In detail, Table 1 recalls the comparison between the value of the skewness 
\bea
\ds H_3(0)= \frac{\langle(\partial u_r/\partial r)^3 \rangle}{\langle (\partial u_r/\partial r)^2\rangle^{3/2}}
\nonumber
\eea 
calculated with the proposed analysis and those obtained by the aforementioned works. It results that the maximum absolute difference between the proposed value and the other ones results to be less than 10 $\%$. 
Furthermore, other studies \cite{deDivitiis_3, deDivitiis_4, deDivitiis_5} have shown that the closure formulas referable to this analysis provide kinetic energy and temperature spectra which exhibit scaling laws in agreement to the theoretical arguments of Kolmogorov, Obukov--Corrsin and Batchelor \cite{Obukhov, Batchelor_2, Batchelor_3}.
Therefore, the adopted hypothesis $\cal H$=max and the consequent distribution (\ref{Pl}) seem to be adequate assumptions at least for what concerns the estimation of turbulent energy cascade main effects.

\bigskip

\section{Results and Discussions}

In order to justify the plausibility of the previous hypotheses, this section presents one statistical analysis of numerical simulations of a simple differential system representing incompressible fluid kinematics. This system is properly chosen in such a way that it can exhibit simple mathematical structure, and chaotic behavior corresponding to an expected high value of entropy $\cal H$.
To achieve this latter condition, we adopt an adequate differential system which shows a "weak" or "reduced" link between velocity and spatial coordinates and an expected huge number of bifurcations per unit time (bifurcations rate).
For this reason, each velocity component is assumed to be depending only on one single spatial coordinate with opportune scaling factors. Thus, the chosen differential system is given by the following equations 
\bea
\begin{array}{l@{\hspace{-0.cm}}l}
\ds \frac{d x}{dt} = u = \sin z, \ \ \ x(0) =  x_0 \\\\
\ds \frac{d y}{dt} = v = \frac{\sin q x}{q}, \ \ \ y(0) =  y_0 \\\\
\ds \frac{d z}{dt} = w = \frac{\sin q^2 y}{q^2}, \ \ \ z(0) =  z_0.
\end{array}
\label{ABC}
\eea
where $q$, giving different scaling factors along the coordinate directions, will be
properly selected to study the system behavior and to obtain a high bifurcations rate.
The velocity field is periodic and ${\cal C} \equiv (0, 2 \pi /q) \times (0, 2 \pi /q^2) \times (0, 2 \pi)$ represents the smallest regions of periodicity.
The velocity gradient is then
\bea
\ds \nabla_{\bf x} {\bf u} =
\left[\begin{array}{ccc}
\hspace{-1.0mm} \ds 0        & 0 &  \cos z \\\\
\hspace{-1.0mm} \ds  \cos q x & 0 &    0   \\\\
\hspace{-1.0mm} \ds 0         &  \cos q^2y & 0   
\end{array}\right]
\eea
and its determinant, $\det({\nabla_{\bf x} {\bf u}}) = \cos q x \ \cos q^2y \ \cos z$,  vanishes in those points where at least one of $q x$,  $q^2y$ and $z$ assumes values $\ds \pi (1/2 \pm k), \ k= 0, 1, 2,...$. Accordingly, large values of bifurcations rate are expected when $q$ is opportunely high.
Observe that, due to its peculiar analytical structure, the system generates trajectories which can differ from isotropic turbulence, while it is not sure that $\cal H$ assumes its maximim value. Nevertheless, as its Jacobian determinant vanishes in numerous points, it is reasonable that the proposed system shows a high entropy and a behavior which can be in some way compared to the fully developed chaos. 
\begin{table}[h]
\centering
\caption{Statistical parameters for $\vert {\bfxi} \vert = 10^{-7}$.
}
\vspace{2. mm}
\begin{tabular}{cccccc} 
q  &  $\bar{\lambda}$  &   $I_1$ & $I_2$ & $\bar{\lambda}/\lambda_+$ & $\sigma/\bar{\lambda}$   \\
\hline 
 2.75    &  0.1879  & 0.9095  &  0.4210   &  0.5339  &  1.7426  \\
 3.75    &  0.2067  & 0.9483  &  0.3726   &  0.5823  &  1.5411  \\
 4.50    &  0.1739  & 0.9003  &  0.4345   &  0.5224  &  1.7981  \\
 17.00   &  0.04525  & 0.9823  &  0.5360   &  0.4859  &  1.7472  \\
 17.50   &  0.04805  & 0.9747  &  0.5197   &  0.4945  &  1.8267  \\
 19.25   &  0.04220  & 0.9897  &  0.5356   &  0.4988  &  1.6693  \\
 20.75   &  0.03970  & 0.9839  &  0.5077   &  0.4994  &  1.7948  \\
 24.00   &  0.03262  & 0.9910  &  0.5593   &  0.4785  &  1.7630  \\
\hline
\hline
\label{table2}
 \end{tabular}
\centering
\caption{Statistical parameters for $\vert {\bfxi} \vert =10^{-3}$.
}
\vspace{2. mm}
\begin{tabular}{cccccc} 
q  &  $\bar{\lambda}$  &   $I_1$ & $I_2$ & $\bar{\lambda}/\lambda_+$ & $\sigma/\bar{\lambda}$   \\
\hline 
 2.75   &  0.1945   & 0.9173  & 0.3998  & 0.5473  &  1.7007  \\
 3.75   &  0.2026   & 0.9442  & 0.3822  & 0.5739  &  1.5740  \\
 4.50   &  0.1685   & 0.8912  & 0.4509  & 0.5081  &  1.8610  \\
 17.00  &  0.04804  & 0.9067  & 0.5103  & 0.4542  &  2.3244  \\
 17.50  &  0.05202  & 0.9358  & 0.4816  & 0.4983  &  2.0231  \\
 19.25  &  0.04448  & 0.9200  & 0.5344  & 0.4555  &  2.2818  \\
 20.75  &  0.04111  & 0.9168  & 0.5150  & 0.4534  &  2.3871  \\
 24.00  &  0.03466  & 0.9176  & 0.5299  & 0.4475  &  2.4426  \\
\hline
\hline
\label{table3}
 \end{tabular}
\end{table} 
Hence, although the regime $\cal H$= max and Eq. (\ref{ABC}) can correspond to different PDFs, say $P_{\lambda}$ and $P_{\lambda  E}$, respectively, it is reasonable that the two distributions show elements in common, expecially for what concerns the interval where $\tilde{\lambda}$ ranges.
In particular, according to this analysis, we would expect  
\bea
\begin{array}{l@{\hspace{-0.cm}}l}
\ds I_1 = \int_{-\lambda_{S}/2}^{\lambda_{S}} {P}_{\lambda  E} d \tilde{\lambda} \lesssim  1,  \\\\
\ds I_2 = \frac{ \ds  {\int_{-\infty}^0 {P}_{\lambda  E} d \tilde{\lambda}} }
{\ds  {\int_0^{\infty} {P}_{\lambda  E} d \tilde{\lambda}}}  \approx \frac{1}{2}, \\\\
\ds \frac{\bar{\lambda}}{\lambda_+} \approx  \frac{1}{2}, \\\\
\ds  \frac{\sigma}{\bar{\lambda}} \approx \sqrt{3}
\end{array}
\label{I1_I2} 
\eea
where $I_1$, $I_2$, $\bar{\lambda}/\lambda_+$ and $\sigma/\bar{\lambda}$ are statistical
parameters related to the PDF shape.
Following Eqs. (\ref{I1_I2}), it is plausible that the occurrences $\tilde{\lambda}>0$ are about two time those $\tilde{\lambda}<0$, and that most of the occurrences of $\tilde{\lambda}$ 
happen in $(-\lambda_{S}/2, \lambda_{S})$, where ${\lambda}_{S}$ is here obtained in function of $\left\langle \tilde{\lambda} \right\rangle_{E}$ by means of Eqs. (\ref{back}) and (\ref{lambda^2}), i.e.
\bea
\ds \lambda_{S} = 4 \left\langle \tilde{\lambda} \right\rangle_{E}
\label{ls}
\eea
being $\left\langle \tilde{\lambda} \right\rangle_{E}$ the average of $\tilde{\lambda}$ calculated through $P_{\lambda E}$. Moreover, $\lambda_+$ and $\sigma$ are expected to be proportional to $\bar{\lambda}$, with $\bar{\lambda}/\lambda_+$ and $\sigma/\bar{\lambda}$ satisfying Eqs. (\ref{I1_I2}).

\begin{figure}[h]
\centering
\includegraphics[width=10.0cm, height=12.2cm]{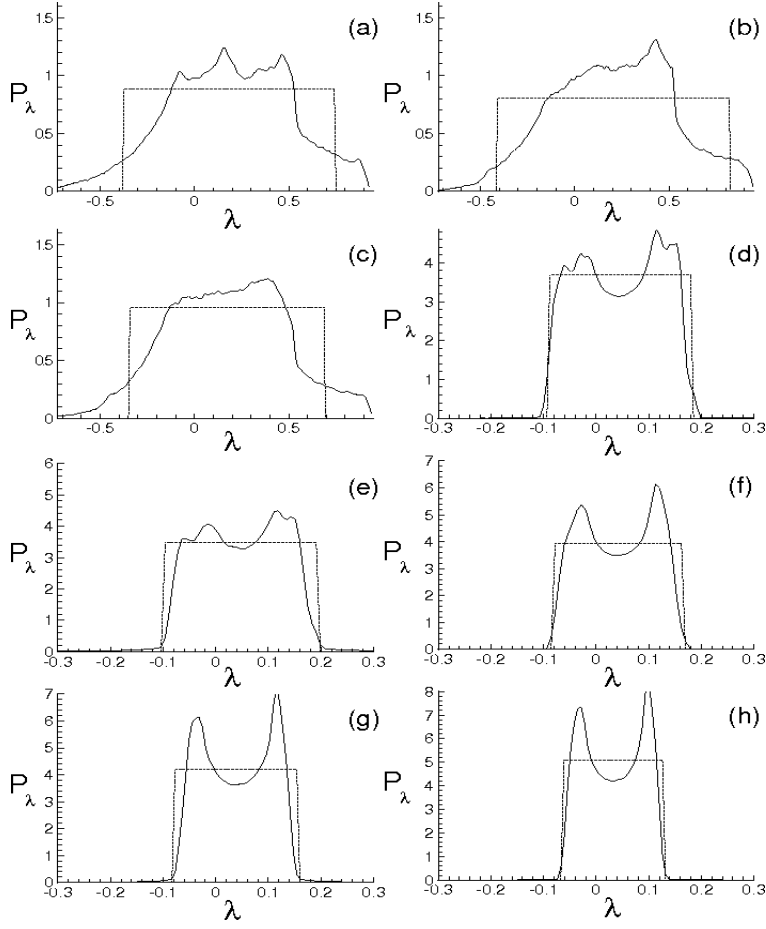}
\caption{Distribution function of $\tilde{\lambda}$ for $\vert {\bfxi} \vert =10^{-7}$.
{\bf (a)}  $q$=2.75, \ 
{\bf (b)}  $q$=3.75, \ 
{\bf (c)}  $q$=4.5, \ 
{\bf (d)}  $q$=17, \ 
{\bf (e)}  $q$=17.5, \
{\bf (f)}  $q$=19.25, \
{\bf (g)}  $q$=20.75, \
{\bf (h)}  $q$=24. \
}
\label{figura_1}
\end{figure}

\begin{figure}[h]
\centering
\includegraphics[width=10.0cm, height=12.2cm]{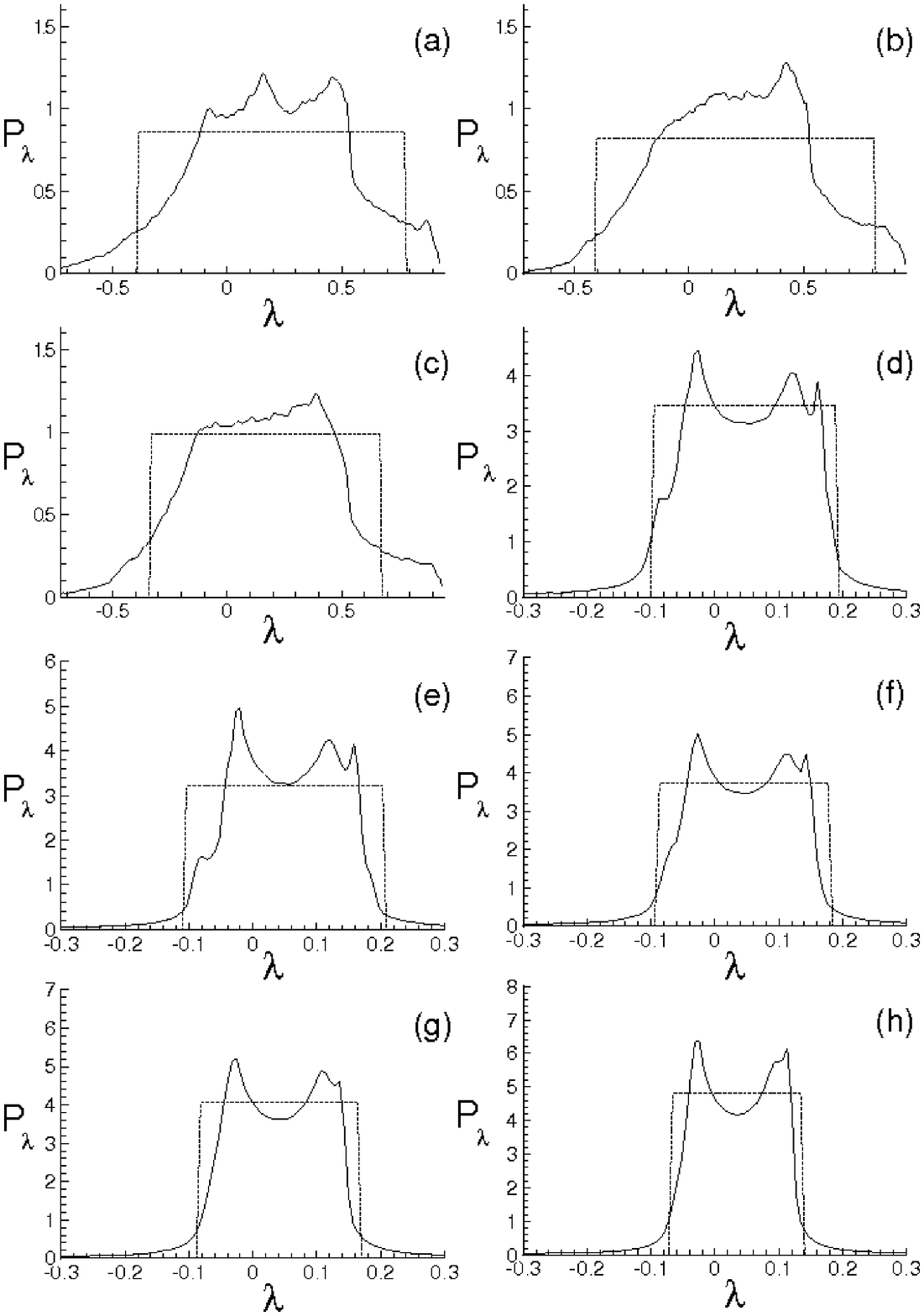}
\caption{Distribution function of $\tilde{\lambda}$ for $\vert {\bfxi}\vert = 10^{-3}$. 
{\bf (a)}  $q$=2.75, \ 
{\bf (b)}  $q$=3.75, \ 
{\bf (c)}  $q$=4.5, \ 
{\bf (d)}  $q$=17, \ 
{\bf (e)}  $q$=17.5, \
{\bf (f)}  $q$=19.25, \
{\bf (g)}  $q$=20.75, \
{\bf (h)}  $q$=24. \
}
\label{figura_2}
\end{figure}

The results, here presented by two sets of eight cases, are obtained by means of numerical simulations. Tables \ref{table2} and \ref{table3} report the mentioned statistic parameters, and Figs. \ref{figura_1} and  \ref{figura_2} show the distribution functions for different values of $q$.
The set of differential equations is given by Eqs. (\ref{ABC}) and by the corresponding
evolution equations of ${\bfxi}$ following Eqs. (\ref{1}).
During the process of integration, the Lyapunov vectors are continuously rescaled
(at each time step) in order to maintain the initial scale $\vert \bfxi(0) \vert$ at which $\bfxi$ and  $\tilde{\lambda}$ are both referred.
The scheme of integration is a fourth--order Runge--Kutta method with automatic adaptive step size, which returns equidistant samples ${\bf x}(t_k)$, ${\bfxi}(t_k)$, $t_k = k \Delta t$, k=1, 2, ..., N, being $\Delta t$ properly computed. Specifically, the time of integration is assumed $T =$ 30000/$\tilde{\lambda}_{0 max}$, being $\tilde{\lambda}_{0 max}$ the maximum Lyapunov exponent at $t=0$, and N= 2. 10$^6$.
Thereafter, the distribution functions of $\tilde{\lambda}$ are numerically computed through statistical elaboration of the simulations data.
In order to obtain an expected high value of the bifurcations rate, 
the assumed values of $q$ are such that $q \in (2, 25)$.
For what concerns the initial conditions, all the simulations are computed for 
$x_0 =0.1$, $y_0=0.2$, $z_0=0.3$, with an initial orientation of $\bfxi$ corresponding to the minimum rising rate of $\rho$.

In line with the chaotic trajectories behaviors, we selected the simulations where $\bf x$ describes almost completely at least $\cal C$, or at least the union of regions of space equivalent to $\cal C$.
The first set of simulations, obtained for $\vert \bf \xi \vert$= 10$^{-7}$, is reported in Fig. \ref{figura_1}, and the other one, depicted in Fig. \ref{figura_2}, is computed for $\vert \bf \xi \vert$ =10$^{-3}$.
Solid and dotted lines represent, respectively, $P_{\lambda E}$ and the estimates of $P_{\lambda}$ obtained from  Eqs. (\ref{Pl}) and (\ref{ls}).
It is apparent that, although the two PDFs differ with each other, the computed values of $\tilde{\lambda}$ are unsymmetrically distributed in such a way that $I_1$, $I_2$, standard deviation and exponents ratio, satisfy the relations (\ref{I1_I2}). In particular, $\sigma$ results to be proportional to $\bar{\lambda}$ in any case with a proportionality constant around to $\sqrt{3}$  (0.8 $\div$ 2.5).

We conclude this section by observing that variations of $q$ and of initial conditions in terms of position ${\bf x}(0)$ and/or orientation of ${\bfxi}(0)$, can produce changing in the shape of $P_{\lambda E}$ and of the aforementioned parameters, depending on how the system describes its phase space. In particular, if $\bf x$ sweeps almost completely at least $\cal C$ --or at least the union of regions of $\left\lbrace \bf x \right\rbrace$ equivalent to $\cal C$-- $\tilde{\lambda}$ is found to be unsymmetrically distributed, and the values of $I_1$, $I_2$, $\sigma/\bar{\lambda}$ and $\bar{\lambda}/\lambda_+$ agree with those given in Eqs. (\ref{I1_I2}).
On the contrary, in the cases where $\bf x$ only partially sweeps $\cal C$ --or the equivalent union of  parts of space-- sizable differences can be observed with respect to the present results in terms of PDF shape and statistical parameters.

\bigskip

\section{Conclusions}

The distribution function of the finite scale local Lyapunov exponent of the kinematic field was studied in homogeneous isotropic turbulence. Based on reasonable assumptions regarding the fully developed chaos and the fluid incompressibility, the shape of such distribution and the range of variations of $\tilde{\lambda}$ are determined. This distribution results to be an uniform function in a proper non--symmetric interval of variations. The results arising from such PDF, in particular the link between $\lambda_+$ and $\bar{\lambda}$ and the closure for the longitudinal velocity correlation equation, agree with those presented in Ref. \cite{deDivitiis_1}, and this should support the hypothesis $\cal{H}$ = max. An alternative way to determine the link between such exponents is also presented, which is based on the alignment property of the Lyapunov vectors. Direct simulations of a very simple differential system representing incompressible fluid kinematics give results which corroborate the hypotheses of the present analysis.

\bigskip

\section{\bf Appendix}

This appendix reports some of the results dealing with the closure of the von K\'arm\'an-Howarth equation, obtained in Refs. \cite{deDivitiis_1} and  \cite{deDivitiis_4}.

For fully developed isotropic homogeneous turbulence, the longitudinal velocity correlation
function 
\bea
f(r) = \frac{\langle u_r({\bf x}) u_r({\bf x}+ {\bf r}) \rangle}{u^2}
\eea
obeys to the von K\'arm\'an-Howarth equation \cite{Karman38}
\bea
\ds \frac{\partial f}{\partial t} = 
\ds  \frac{K(r)}{u^2} +
\ds 2 \nu  \left(  \frac{\partial^2 f} {\partial r^2} +
\ds \frac{4}{r} \frac{\partial f}{\partial r}  \right) +\frac{10 \nu}{\lambda_T^2} f 
\label{vk-h}
\eea
the boundary conditions of which are
\bea
\begin{array}{l@{\hspace{+0.2cm}}l}
\ds f(0) = 1,  \\\\
\ds \lim_{r \rightarrow \infty} f (r) = 0
\end{array}
\label{bc0}
\eea
where $\nu$ is the fluid kinematic viscosity, and $u \equiv \sqrt{\langle u_r^2({\bf x}) \rangle}$ 
follows the kinetic energy equation obtained from Eq. (\ref{vk-h}) for $r\rightarrow 0$ \cite{Karman38}
\bea
\ds \frac{d u^2}{d t} = - \frac{10 \nu}{\lambda_T^2} u^2 
\eea
being $\lambda_T \equiv \sqrt{-1/f''(0)}$  the Taylor scale.
The quantity $K(r)$ gives the energy cascade, being linked to the 
longitudinal triple velocity correlation function $k$ 
\bea
\begin{array}{l@{\hspace{+0.0cm}}l}
\ds K(r) = u^3 \left( \frac{\partial }{\partial r} + \frac{4}{r} \right) k(r), 
\ \ \mbox{where} \ \ 
\ds k(r) = \frac{\langle u_r^2({\bf x}) u_r({\bf x}+ {\bf r}) \rangle}{u^3}
\end{array}
\eea
Thus, the von K\'arm\'an-Howarth equation gives the relationship between the statistical
moments $\left\langle (\Delta u_r)^2 \right\rangle$ and $\left\langle (\Delta u_r)^3 \right\rangle$ in function of $r$.

The analysis in Refs. \cite{deDivitiis_1} and \cite{deDivitiis_4} provides the
closure of the von K\'arm\'an-Howarth equation, and expresses  
$K(r)$ in terms of longitudinal velocity correlation  
\bea
\ds K(r) = u^3 \sqrt{\frac{1-f}{2}} \frac{\partial f}{\partial r}
\label{K}
\label{K closure}
\eea
The skewness of $\Delta u_r$ is \cite{Batchelor53}
\bea
\ds H_3(r) \equiv 
\frac{\langle (\Delta u_r)^3 \rangle }{\langle (\Delta u_r)^2 \rangle^{3/2}} 
=
\frac{6 k(r)}{(2(1-f(r)))^{3/2}}
\label{H3}
\eea
Therefore, the skewness of $\partial u_r/ \partial r$ is 
\bea
H_3(0) = -\frac{3}{7}
\eea

\bigskip

\section{Competing Interests}

The author declares that there is no conflict of interests regarding the publication of this article.

\bigskip

\section{Acknowledgments}

This work was partially supported by the Italian Ministry for the Universities 
and Scientific and Technological Research (MIUR), and received no specific grant from any funding agency in the public, commercial or not-for-profit sectors.

\bigskip

\bigskip

\end{document}